# Credential Masquerading and OpenSSL Spy: Exploring ROS 2 using DDS security

Vincenzo DiLuoffo, William R.Michalson, Berk Sunar

*Robotics, Electrical and Computer Engineering*
*Worcester Polytechnic Institute (WPI)*
*100 Institute Rd, Worcester, MA 01609*
*{vdiluoffo, sunar, wrm } @wpi.edu*

*Abstract*— The trend toward autonomous robot deployments is on an upward growth curve. These robots are undertaking new tasks and are being integrated into society. Examples of this trend are autonomous vehicles, humanoids, and eldercare. The movement from factory floors to streets and homes has also increased the number of vulnerabilities that adversaries can utilize. To improve security, Robot Operating System (ROS) 2 has standardized on using Data Distributed Services (DDS) as the messaging layer, which supports a security standard for protecting messages between parties with access control enforcement. DDS security is dependent on the OpenSSL and a security configuration file that specifies sensitive data location. DSS Security assumes that the underlining Operating System (OS) is secure and that the dependencies are consistent, but ongoing integrity checks are not performed. This paper looks at two vulnerabilities that we exploit using an OpenSSL spy process and a security property file manipulation. An overview of each exploit is provided with an evaluation of mitigation technologies that may be employed in client computers, servers, and other areas. Since, ROS 2 and DDS run in user space, these processes are prone to vulnerabilities. We provide recommendations about mitigation technology, as currently autonomous platforms are being deployed without safe-guards for on or off-line threats. The Trust Platform Module (TPM) is new to robotic systems, but the standard usage model does not provide risk mitigation above the OS layer for the types of attacks we discuss.

## I.    INTRODUCTION

Modern robots are constructed with sensors, controllers, communications, motors, hardware accelerators as well as software forming a cognitive layer for processing and controlling the robot. Autonomous robots are often fully autonomous, putting their software and hardware all in one location, providing an adversary with a complete system with little, if any, physical security.

This makes physical attacks on robots much easier than attacks on corporate managed computers, since systems are typically under system management and are physically protected by the building that houses them. As robots move from factory floors into society this physical protection is removed making systems more vulnerable.

To address security in robotic systems, ROS 2 with DDS Security allows online data in-motion encryption with access control protection. DDS security is dependent on the OpenSSL and a security configuration file that specifies sensitive data location. DSS Security assumes that the underlining Operating System (OS) is secure and that the dependencies are consistent, but ongoing integrity checks are not performed. However, off-line and on-line exploits can involve software or hardware attacks, especially when robots are out in the wild. Research is in the early stages of investigating autonomous vehicle security [1] [2], artificial intelligence and robotics [3] while others are looking at the performance related to security [4] and creating isolation containers from memory restrictions using ARM Trustzone [5]. However, these approaches tend to focus on individual security threats and ignore viewing the threat environment as a whole; that is, taking what we refer to as a holistic approach to autonomous robot security.  An example vulnerability analysis can be found in [3].

Operating systems are generally known for being susceptible to various types of exploits; a common one being the exploitation of privilege escalation [6] [7] [8] to gain access. The use of secure and trusted boot code mitigates some of these potential threats, but only up to the OS layer.  In Linux, the OpenSSL library executes in the user space, so the application space is vulnerable to many exploits. These exploits include buffer overflow, timing attacks, and injected malware, just to name a few.

This paper identifies exploits in the user/ application space related to OpenSSL and the property configuration file for the ROS 2/DDS security usage model. Serval technologies are examined that potentially mitigate these vulnerabilities.

We describe four different scenarios where these types of attacks affect the behavior of the system.

- *In use case one, we have a swarm of drones that are performing a surveillance mission around a building. Each night at the same time the drones are dispatched. We will call this the spy intercept scenario. The drones have been infected by the adversary's altered OpenSSL library and now the data is being streamed to a remote server where all the data is being dumped. The adversary now has the capability to review the mission, data captured from the drones and can determine where potential schedule gaps of surveillance exists.*

- *In use case two, an adversary has placed an altered OpenSSL library on the platform and is able to siphon data from a sensor. In this example, a camera mounted on the robot can send data to a remote server for the adversary to view, this is called stealing services.*

- *In use case three, an autonomous vehicle can be repurposed. An adversary has altered the OpenSSL library on a vehicle and now has access to the keys. This means that control of the vehicle can be achieved for late night runs when owner is asleep. The car can be returned without being noticed.*

- *In use case four, the configuration file is altered by the adversary to change the credentials in the property file, this will enable the adversary to take control of the robot platform without being noticed. The property configuration file and OpenSSL library manipulation can occur on the same platform to provide additional control to the adversary.*

The above use cases are a set of examples of OpenSSL or property file exploits. The list of exploits can grow and will enable an adversary to have control of an autonomous platform.

## II. SUMMARY OF RESULTS

### A. Exploiting the usage of OpenSSL

In this paper, we show that a compromised OpenSSL library can intercept sensitive information while victim participants believe they are sending secure publisher and subscriber messages. In ROS2/DDS, security on a Linux system is enabled by having an interface between the vendor's security plugins and the OpenSSL library. The governance and participant policies define the security behavior within the domain. These behaviors also define the protection kinds to be used and therefore, the cryptographic algorithms associated with any communications. The vendor's security plugin implementation is proprietary, and how or what functions are being called from within the OpenSSL library is also unknown. However, by replacing the original OpenSSL library with an altered "spy" OpenSSL library, the interactions between participants become visible. Using this approach, we demonstrate that information can be captured on a victim machine. While the demonstration presented here was done locally, a remote server can be utilized where all information is sent to it for post processing or data manipulation, as in a man in middle attack. As autonomous robots tend to have many, if not all, of their nodes on a single platform this type of exploit can be a severe threat. To thwart this type of attack, we identify mitigation techniques that may be applied for detecting unknown spy processes in the Linux user space software stack.

### B. Manipulating the configuration file for misuse

Another exploit we explore is the misuse of the ROS 2/DDS Security property file where security credentials are configured. This attack involves an adversary manipulate the property data using masquerading credentials. A single, unprotected, configuration file supports the credentials of the CA's Certificate (authentication and document sign); participant's certificate and private key; as well as the signed governance and participant's policies. These configuration files point to locations on the file system where the sensitive data is located, and changes to these files are undetected; meaning that an adversary can substitute their own credentials. Depending on the CA's issuing policy, an email can be the simplest form of authentication for issuance, so an adversary can get their own credentials from the initial CA versus just replacing the configuration file with adversary data. The private keys are base 64 encoded using a privacy enhanced mail (PEM) format [9], depending on the security policy these can be password protected. However, there is no difference between how the security plugin checks for an illegal set of configuration parameters versus a legitimate set; therefore, an adversary can change the credential parameters without detection.

## III. BACKGROUND

The Robot Operating System (ROS) 2 uses the Data Distribution Service (DDS) [10] as the transport layer. An extension to DDS is the DDS security standard, which provides the data protection layer on the Real Time Publisher Subscriber (RTPS) [11] data. DDS security [12] is implemented by a vendor's security plugin. The standard specifies five Security Plugin Interfaces (SPIs); Authentication, Access Control, Cryptographic, Logging and Data Tagging.

Authentication and Cryptographic operations are defined in the standard, but the implementation is dependent on the vendor. There are no test vectors or compliance tests for the security plugin implementation. The standard simply defines specific algorithms for authentication and cryptographic operations - OpenSSL [13] is a common cryptographic library for providing these algorithms.

When ROS2 is built on top of a DDS implementation it must have all its dependent paths configured. The build process must include a path to OpenSSL or another, cryptographic library. In the case of using RTI 5.3 DDS with support for DDS security on Linux, the security plugin is configured to use OpenSSL native.

An XML or YAML (YAML Ain't Markup Language) structured file is used to define the parameters for the security plugin. These parameters are the Certificate Authorities public keys, for identity and policies, signed governance and permission policy files, and the participant's certificate and private key. The file path for each of the parameters is also set in the property configuration file. Each of the policies (governance and permission) enables a protection kind on the message data. Available protection kinds are discovery, liveliness, RTPS, metadata and data. An explanation of these protection kinds can be found in [3] [12]. The policies also define the access control for the domain; that includes the nodes, and who has write or read authorization.

To use a DDS implementation with ROS 2, the environment is setup by configuring the paths for the DDS implementation; on Linux this is done in the users bashrc file. The DDS configuration includes the path to the OpenSSL library. Next, the DDS environment and ROS 2 are compiled with the DDS. Since, the OpenSSL library can be updated by upstream providers for Linux, this library lives in user space. In most cases the OpenSSL library is used as a dynamic shared library, so that changes to the OpenSSL library are independent of the application. In the case where the OpenSSL library is a static shared library both DDS and ROS 2 must be recompiled to take advantage of any new patches or updates.

How each participant discovers, and shares information is enforced by the Governance and the Participant security policies. The transfer of data between two participants in the global data space is performed using the Real Time Publish Subscribe protocol (RTPS) [11]. The use of QoS profiles allows the RTPS communication layer to provide reliability and support for realtime environments where critical processes are under time constraints to complete. QoS profiles or property files also enable the security parameters for the DDS Security deployment model. When security controls are turned on, discovery (who is publishing, sequence numbering and in-line QoS), reliability (heartbeat, ack, nack) metadata and payload data can be encrypted providing the highest security protections. The usage of the protection kinds corresponds to what types of cryptographic algorithms are called. In order to exchange data between participants in a secure manner, each participant is initially issued an identity certificate using a public key algorithm. The authentication plugin performs the initial setup for key exchange and that is followed by security controls performed by the cryptographic plugin. The following set of cryptographic equations are given as reference (DDS Security standard) and are discussed in the attack overview section.

For the authentication plugin, a participant is issued a certificate based on one of the following types of algorithm/key definitions, RSASSA-PSS 2048 or ECDSA 256 bits.

RSASSA-PSS is defined as [14]:

$$n = p * q \qquad (1)$$

Where n = is the modulus, p and q = are prime numbers, and d = is the private exponent. The modulus, n, specifies the key length in bits. The message m, is encrypted using the private key to produce a digital signature, s.

$$RSASSA - PSS - SIGN = ((n, d), m) \qquad (1a)$$

To verify the digital signature, the public key, e = is the public exponent is used to decrypt the message.

$$RSASSA - PSS - VERIFY = ((n, e), m, s) \quad (1b)$$

ECDSA is defined as [15]:

$$Qa(x, y) = d * G \qquad (2)$$

where d = private key, G = field of points, and Q (x, y) = public key curve point

Sign operation:

$$r = x \bmod n \qquad (2a)$$

where r is part of the signature pair, n = integer order of G

$$s = k^{-1}(z - rd) \qquad (2b)$$

where k= random integer, z = left most bit of the hash

Verify operation

$$w = s^{-1} \bmod n \qquad (2c)$$

$$u_1 = z \cdot w \bmod n \text{ and } u_2 = r \cdot w \bmod n \qquad (2d)$$

$$(x, y) = u_1 \times G + u_2 x \, Qa \qquad (2e)$$

where r = x mod n

Key Agreement is used to exchange symmetric keys using public keys. The DH key is 2048-bits MODP Group with 256-bits Prime or ECDH + prime 256 v1 as stated in the standard [12]. These are considered ephemeral keys, that are temporary and only for the session.

ECDH is defined as [16]:

$$(x_k, y_k) = d_a Q_b \qquad (3)$$

where $Q_b$ = public key of user2

$$(x_k, y_k) = d_b Q_a \qquad (3a)$$

where $Q_a$ = public key of user1

$x_k$= shared secret between user1 and user2 is used to exchange symmetric keys.

For the cryptographic plugin, AES_GCM and AES_GMAC are used for authenticated encryption and decryption functions, that are symmetric key operations. Symmetric key operations are low latency, especially when cryptographic modes are combined into an atomic operation. AES_GCM is mostly discussed in the papers, but GMAC is a mode of GCM in which no plain text is supplied and the output is the authenticated field. GCM -128 and GCM -256-bit keys are specified in the standard.

Authenticated Encryption is defined as [17]:

$$C = P \oplus MSB(E(K, Y)) \qquad (4)$$

where C = ciphertext, P = plaintext, MSB = Most Significant Bit, E = encryption of Y using K = key

$$T = MSB(GHASH(H, A, C) \oplus E(K, Y)) \qquad (4a)$$

where T = tag, H = hash, A = additional authenticated data, C= ciphertext, IV = nonce

Authenticated Decryption is defined as [17]:

$$T = MSB(GHASH(H, A, C) \oplus E(K, Y)) \qquad (4b)$$
$$P = C \oplus MSB(E(K, Y)) \qquad (4c)$$

A hash function is called the Secure Hash Algorithm (SHA - 2). Hash functions are used to protect the integrity of the data from being altered. Using a Hash Message Authentication Key Code HMAC [11] falls under this category also, except that it uses a key as part of the hash algorithm.

SHA-2 is defined as [18]:

$$SHA\text{-}2 = (M) \qquad (5)$$

HMAC is defined as [19]:

$$HMAC = (K, M) \qquad (6)$$

The OpenSSL library has two functions, one is for the cryptographic algorithm operations and certificate support, the second is for client and server support for the secure socket layer / transport secure layer. The OpenSSL command line executable is in the `/usr/bin` directory on Ubuntu Linux. The directory `/usr/lib/ssl` points to `/etc/ssl/openssl.cnf`, which defines how certificates are created. Under the ssl directory are the certs, misc, and private directories. The certs directory has several certificates, the misc directory handles the certificate generation and private is owned by root as the key store. The header files are in the `/usr/include/openssl` directory and the shared, libssl and libcryto libraries are in the `/lib/x86_64-linux_gnu` directory. The OpenSSL software layers are illustrated in Figure 1, starting with a set of abstracted higher level APIs, followed by the lower level APIs and supporting utilities and the hardware interface being at the lowest [20].

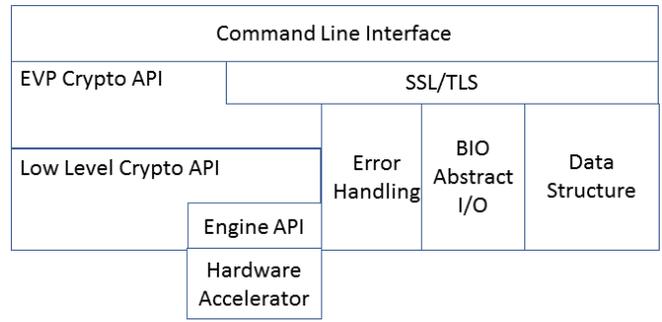

*Figure 1: OpenSSL Architecture*

The RTPS protocol standard defines the message structure for data exchange and the DDS security standard defines the new RTPS message wrappers that conform to the RTPS messages[12]. The wrappers provide the protection on the message structure using encryption, message authentication and/or digital signatures.

When security is enabled the RTPS messages are transformed with special wrappers and still conform to the protocol standard. Figure 2 [12] shows a regular message stack on the left and the secure transformation on the right. Depending on the protection kind being specified the secure wrappers can be applied at the RTPS message, meta sub-message and/or at the payload sub-message levels.

The *rtps_protection_kind* provides protection on the entire message including the message header. Instead of protecting the entire message, a finer control can be achieved using the *metadata_protection _kind* and *data_protection_kind* as two independent operations.
The *metadata_protection_kind* provides protection on the sub-message header and the sub-message elements that includes the GuidPrefix, EntityId, SequenceNumber, SequenceNumberSet, FragmentNumber, ragmentNumberSet, VendorId, ProtocolVersion, LocatorList, Timestamp, Count, and ParameterList elements. The *data_ protection_kind* is only protecting the serialized payload, so depending on the security requirements different levels of protection can be achieved using the Governance policy.

As shown in Figure 2, the SRTPS_PREFIX is used to wrap a complete RTPS message and the sub-messageId is set to 0x33. This is followed by SEC_PREFIX, that is used to wrap a RTPS sub-message and the sub-messageId is set to 0x31. The SecurePayload has a sub-messageId set to the value 0x30. The counterparts to the prefixes are the postfix messages that provide a method to validate the authenticity of the RTPS sub-messages. The SEC_POSTFIX has a sub-meeasgeId of 0x32 and the SRTPS_POSTIX has the sub-messageId of 0x34. The SecureDataHeader contains the cryptographic transformation information and is followed by the information that authenticates the result from the cryptographic transform.

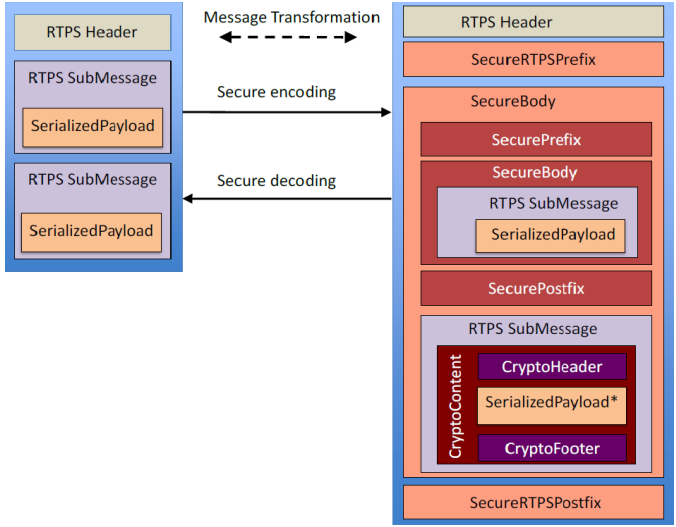

*Figure 2: Secure Transform of RTPS messages*

| Mapping Name | Name | Value |
|---|---|---|
| Entity Ids Mapping | SEDPbuiltinPublicationSecureWriter | {{ff,00,03}, c2} |
| | SEDPbuiltinPublicationSecureReader | {{ff,00,03}, c7} |
| | SEDPbuiltinSubscriptionSecureWriter | {{ff,00,04}, c2} |
| | SEDPbuiltinSubscriptionSecureReader | {{ff,00,04}, c7} |
| | BuiltinParticipantMessageSecureWriter | {{ff.20,00}, c2} |
| | BuiltinParticipantMessageSecureReader | {{ff.20,00}, c7} |
| | BuiltinParticipantStatelessMessageWriter | {{00.20,01}, c3} |
| | BuiltinParticipantStatelessMessageReader | {{00.20,01}, c4} |
| | BuiltinParticipantVolatileMessageSecureWriter | {{ff,02,02}, c3} |
| | BuiltinParticipantVolatileMessageSecureReader | {{ff,02,02}, c4} |
| | SPDPbuiltinParticipantsSecureWriter | {{ff, 01,01}, c2} |
| | SPDPbuiltinParticipantsSecureReader | {{ff,01 01}, c7} |
| | | |
| Member | PID_IDENTITY_TOKEN | 0x1001 |
| | PID_PERMISSION_TOKEN | 0x1002 |
| | PID_PARTICIPANT_SECURITY_INFO | 0x1005 |
| | PID_PROPERTY_LIST | 0x0059 |
| | | |
| Cryptographic | CRYPTO_TRANSFORMATION_KIND_NONE | {0, 0, 0, 0} |
| | CRYPTO_TRANSFORMATION_KIND_AES128_GMAC | {0, 0, 0, 1} |
| | CRYPTO_TRANSFORMATION_KIND_AES128_GCM | {0, 0, 0, 2} |
| | CRYPTO_TRANSFORMATION_KIND_AES256_GMAC | {0, 0, 0, 3} |
| | CRYPTO_TRANSFORMATION_KIND_AES256_GCM | {0, 0, 0, 4} |

The set of defined builtins in the DDS Security standard are used to enable the interoperability between vendors. Section 7 of the DDS security standard defines the mappings of Entity Id values for the secure builtin data writer and data reader. Mappings for builtin participants, CryptoTransformationKind and CryptoTransformKeyId are defined in section 8, and key material is defined in section 9 [12] Table 1 shows the list of partial values that are transmitted during the RTPS message exchange.

## IV. Attack Overview

The five security plugs are provided by the vendor as object code that treats, the OpenSSL library as a black box. Thus, to understand how the security services use the OpenSSL library, we examined the OpenSSL library to see what cryptographic functions were being called. The cryptographic functions are defined in the security standard, but since OpenSSL has two different levels of API as shown in Figure 1, we needed to add dump routines to save data that is normally internal to OpenSSL library operation into files and use that data to identify the use of specific OpenSSL function calls. This process is not trivial as for some cryptographic functions there are initial, update and final sets of API calls to complete a single cryptographic operation.

Our first step was to add these dump routines into the cryptographic functions, specifically gcm128, hmac, e_aes, evp_digest, and ech_key. Since the property file already has the identity key values in the local directory there was no need to add additional dump routines in the code. The dump routine used was the hexdump function found in the OpenSSL library test code but modified for writing to a local file. In most cases the output from the individual routines were modified to match the hexdump function parameters. For example, *static void hexdump (File *f, const char *title, unsigned char * s, int l)* was the modified dump routine and for working with big numbers the format needed to convert was *BN_bn2hex().* The basic function call was *hexdump (stdout, "title", variable, variable length).* This call was placed into each of the above files and within each initial, update and final API function. Figure 3 shows the flow of cryptographic operations performed during authentication between two parties along the horizontal path and how data protection is performed is shown in the vertical path.

As part of the authentication, a common shared secret is established using a key exchange algorithm (equations 3 and 3a), this value is hashed and used within the CryptoKeyFactory, CryptoKeyExchange, CryptoTransform and LogOptions data structures for identity and authentication token exchange (see tables 36, 37, 38 and 39 of [12]). As a final step, the participant will digitally sign the data using his/her private key (all equations in 1 or 2 depending on algorithm and equation 5 for the hash operations). As part of the dump routines this data is captured to reveal the sensitive data, such as shared secret, message token parameters and private keys used to digitally sign the token data.

The vertical path is used to perform integrity and confidentiality data protection. From the hashed data branch to applying an integrity check using HMAC (equation 6), the data is than transformed using AES_GCM (all equation in 4). The key computation and cryptographic transformations formulas (see table 73) in the DDS security standard [12] provides key convolutions and transforms. The dump routines reveal all the data and keys used during these cryptographic operations. The output data from each of the dump routines must be converted from hex to ascii to show the human readable data being exposed. Even though some of the sensitive data is transient, an adversary can still manipulate the data and expose a threat as discussed in the use cases above.

To confirm our data dump routines and modified OpenSSL Library performed as desired, we tested in two different environments. The first test environment was configured using

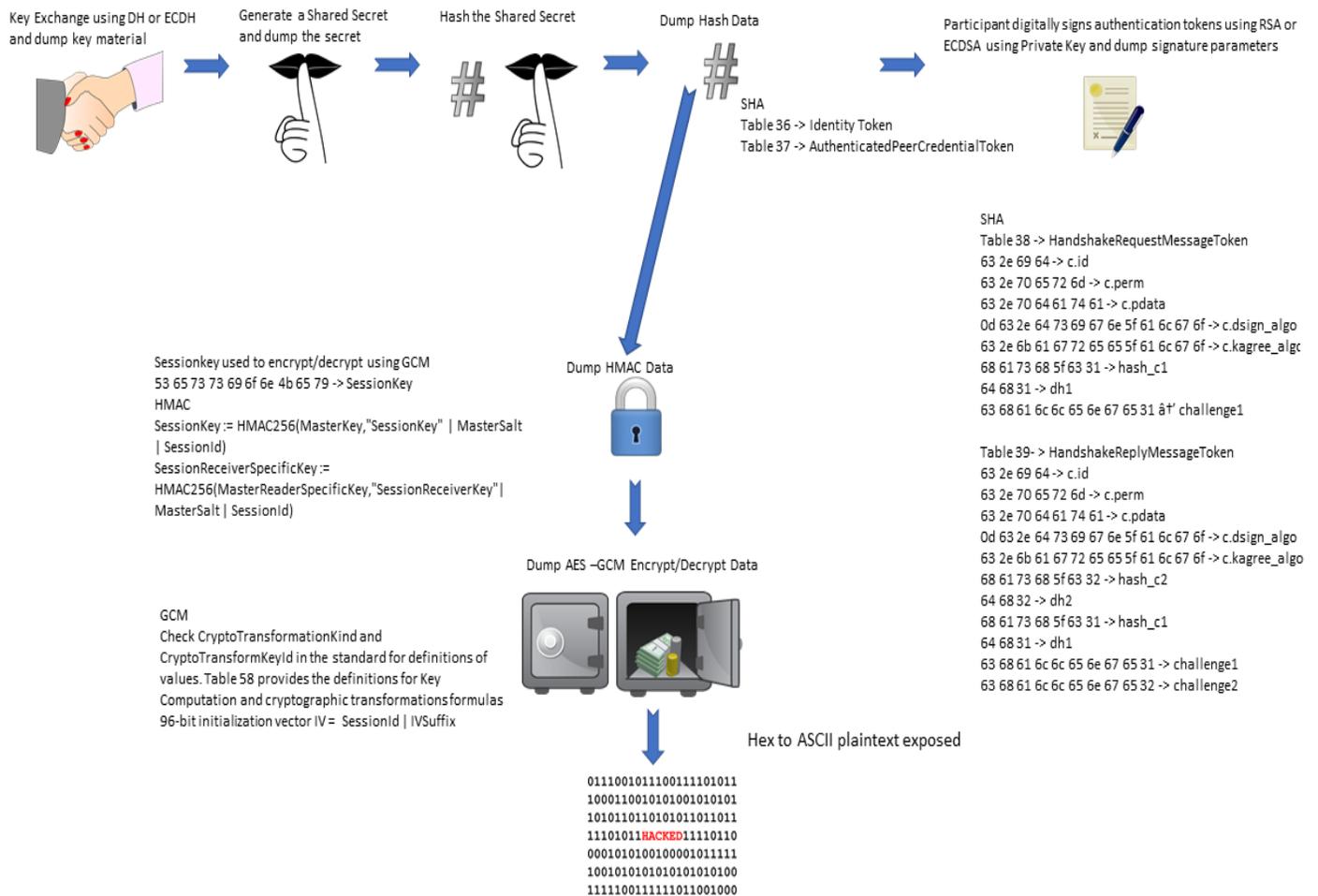

*Figure 3: DDS Security data dump flow*

a Lenovo W541 computer with Linux Ubuntu 16.04. The

OpenSSL library used was 1.0.2m, but newer releases can also be used. The Real-Time Innovation (RTI) 5.3 DDS with security and RTI Perftest 2.4 [21] were used to compile and run the tests. Wireshark 2.4.6 was used to capture the RTPS packet network traffic. The OpenSSL source files were downloaded from OpenSSL.org, modified, and compiled. The files were extracted in the /usr/src/openssl-1.0.2m directory. All the changes were done in the /crypto directory and gcm128 was under the /crypto/modes directory.

To verify that the modified OpenSSL libraries were still working correctly, outputs were compared to the downloaded OpenSSL test vectors. The following commands were used to compile and test:

1) /usr/src/openssl-1.0.2m$ ./configure shared (creates lib* and openssl files)
2) /usr/src/openssl-1.0.2m$ sudo make (compile)
3) /usr/src/openssl-1.0.2m$ sudo make test (test)

Each of the dump routines were checked against the test vectors, or against the standards being followed by the OpenSSL developers. For example, RFC 7027 for the brain pool curve test vectors. The matching of the dump routine to the test vectors gave confidence that the dump routines and formatting were providing the same results from a known good source. Now that the OpenSSL library has been compiled and tested we compiled against the RTIPerftest. For this to occur we first directed the path to our OpenSSL by creating a symlink: ln -s /path/to file /path/to/symlink and by changing the bashrc file RTI_OPENSSLHOME = /usr/src/openssl_1.0.2m. RTIPerftest compiled using: ./build.sh –platform x64Linux3gcc5.4.0 –secure –openssl-home /usr/src/openssl-1.0.2m.

Our next step is to collect data using wireshark and saving only those transactions that used the RTPS protocol only to, reduce the amount of network traffic data collected. We ran the publisher and subscriber using no security to give us a baseline with the following commands in two console terminals:

1) ./bin/x64Linux3gcc5.4.0/release/perftest_cpp -pub -datalen 63000 -executionTime 1
2) ./bin/x64Linux3gcc5.4.0/release/perftest_cpp -sub -datalen 63000

In Figure 4 is the result of no security shown using Wireshark.

*Figure 4: Wireshark plain run of publisher and subscriber transaction*

We use the following commands to run with discovery and liveness protection kind using encryption:

1) ./bin/x64Linux3gcc5.4.0/release/perftest_cpp -pub -datalen 63000 -secureEncryptDiscovery -executionTime 1
2) ./bin/x64Linux3gcc5.4.0/release/perftest_cpp -sub -datalen 63000 -secureEncryptDiscovery

From our dump routines there should be several files that have been created in the specified directory paths, Figure 5 shows the output from Wireshark where the top blue highlight is one of the secure wrappers as shown in Figure 2. The lower blue highlight is the expanded view that represents the encoding of the SecurePayload with a value of 0x30 using a crypto_transformation with AES-GCM 256 that correlates to value {0, 0, 0, 4} as shown in Table 1. The payload is followed by a closing post sub message with a value of x32.

*Figure 5: Wireshark detailed view into RTPS protocol transaction using secureEncryptDiscovery enabled*

Figure 6 shows the dump routine file that was generated showing the gcm key, iv and final tag. These values can be used to decrypt the messages if they aren't already in the clear in another file, like the digest output file.

*Figure 4: Output from dump routine that shows key, iv and final tag*

Figure 7 shows the output from an AES_GCM 256 bit script to decrypt the data using the key, iv and final tag [22]. We show that by obtaining the relevant algorithm parameters that decryption is validated by the auth_ok_ = true and plain text being showed between the brackets in figure 7.

*Figure 5: Output from gcm decrpt script*

In the ecdh file this is the data captured from the key exchange between publisher and subscriber. Figure 8 shows the public; private and shared secret key being dumped.

*Figure 6: ECDH handshake between two participants to reveal the shared secret key*

The second test environment was configured using a Lenovo L450 with Ubuntu 18.04, RTI DDS with security 5.3.1, ROS 2 Bouncy Bolson release and SROS enabled. We used the same OpenSSL Library as in the first test environment, release 1.0.2m that was modified. We copied the files into their respective locations. The results from our data dump routine revealed the plaintext for the Hello World example using SROS as shown in Figure 9, running the talker and listener nodes.

*Figure 7: SROS example of talker and listener*

From our output dump routines, we can exploit the DDS security and view the data being exchanged between two parties. An adversary can easily replace the libcrypto.so.* and libssl.so.* files in the /lib/x86_64-linux_gnu directory using an escalated privilege exploit. The dump routine can be easily extended to write the data to a remote server. Validation was performed using Ubuntu 16.04 on a ThinkPad® model L450, where the files libcrpto.1.0.0 and libssl.1.0.0 were in the /lib/x86_64 directory. These files were replaced with our spy libraries and symlinks were created to libcrypto.so and libssl.so from these two libraries.

    libcrypto.so.1.0.0 -> libcrypto.so
    libssl.so.1.0.0 -> libssl.so

We copied the openssl file into /usr/bin, which provided the configuration needed to capture the data being used by RTI DDS security plugin.

The basic concept of using DDS security is to have all participants use public key and have the policy rules digitally signed. This level of trust is established at the Certificate Authority since both identity and document certificates are issued by a single entity. Depending on the deployment model different CAs can issue identity and document signing operations. The validation in a two-party exchange is performed by having the chain of trust, this means that the issuing CAs public key must be available during a public key validation operation. When security is enabled each participant must have their identity credentials and the signed governance and permission files on the same platform. Since, autonomous systems are self-contained, credentials and policy rules are also stored on the platform. A property file defines all the participants (publisher and subscriber) credentials, identity CA's public key file, and document signed CA's public key, and signed policy rules (governance and permission) located within a directory. An example of a property file is shown in Figure 10 where the credential locations are defined. The security plugin is defined at the top of the file, followed by the CA certificate and PEM data. The domain governance file location is defined followed by the participants (publisher and subscriber). In each of those sections the permission file location is defined and as well as the private key locations. This example is for a dynamic linking security property file and a static method can be used but would need to be compiled into the code.

*Figure 8: Property file example for a publisher and subscriber*

Since changes in credentials and locations will occur, the static method is less flexible since this requires the program to be compiled for each change. In this example the CA's , publisher, subscriber and signed files are known by the security plugins, since these files are parsed for the required information to perform, authentication, authorization and cryptographic operations using the parameters provided.

However, no checks are performed on these files, so that manipulation of the parameters or the files themselves can be achieved by an adversary. An adversary can masquerade the credentials with their own set or change the policy rule files with system parameters / topic names to be self-signed. We see this as a serious vulnerability, since a property file can be altered, and no checks are in place to detect the tampering or credentials being replaced.

## V. EVALUATE MITIGATION OPTIONS

We will evaluate several technologies that might mitigate one of the two vulnerabilities and provide a recommendation for both. The first technology is the Trusted Platform Module, the second being the Security Services for DDS security plugins using ARM TrustedZone and the recommendation is the combination of Integrity Measurement Architecture (IMA)/Extended Verification Module (EVM) with the TPM.

Several industries are using the Trusted Platform Module (TPM), part of the Trusted Computer Group (TCG) to establish root of trust for the PC/Server market. This is being extended into the mobile and Internet of Things as proposed by GlobalPlatform [23] and the Industrial Internet of Things Security Framework [24]. The introduction of using TPMs in robots is still a novel thought. Trusted boot is the process of taking measurements during the boot process from firmware to Operating System (OS) and validating the measurements against a known good set of values by a 3rd party. The TPM is a small microprocessor like a smart card but has a different structure for how hash values are stored in platform configuration registers (PCR). Figure 9 shows the TPM 2.0 structure with support for newer algorithms including Elliptic Curve cryptography and SHA 256 bit.

*Figure 9: Trusted Platform Module 2.0*

The TPM functionality can be implemented in software as well but eliminates the hardware protection features found in some manufactures products. In either case, static or dynamic root of trust measurements are up to the OS executing. In the case of static case PCR (0 to 7) are used in the boot process from Power on Reset (PoR) to OS and in the dynamic case

PCR (17 to 20) when the x86 instruction halts the processor into a known state [25]. Table 1 provides the layout structure for the TMP as shown below [25] [26].

*Table 2: Platform Configuration Register Layout*

| PCR Number | PCR Value |
|------------|-----------|
| 0 | BIOS |
| 1 | BIOS Configuration |
| 2 | Option ROMs |
| 3 | Option ROM configuration |
| 4 | MBR (master boot record) |
| 5 | MBR configuration |
| 6 | State transitions and wake events |
| 7 | Platform manufacturer specific measurements |
| 8 to 9 | Static operating system |
| 10 | Integrity Measurement Architecture (IMA) |
| 11 to 15 | Static operating system |
| 16 | Debug |
| 17 | DRTM and launch control policy |
| 18 | Trusted OS start-up code (MLE) |
| 19 | Trusted OS (for example OS configuration) |
| 20 | Trusted OS (for example OS Kernel and other code) |
| 21 | as defined by the Trusted OS |
| 22 | as defined by the Trusted OS |
| 23 | Application support |
| 24 | |

The purpose of the trusted boot using a TPM was for network admission, meaning that before the computing node was granted access to the network it was validated against Policy Enforcement Point and Policy Decision Point entities. The TMP is used to store the hash values collected during the boot process and then digitally sign a quote (all the hash values), that was sent within a Tunneled Network Connection protocol. The known good values or golden measurements were stored and validated by a 3 rd party verifier, and the result of the compare was sent to the PDP to remediate or allow the compute node to gain access. The remediation process, for example is to update the compute node with the latest software patches or image. Once updated the compute node would validate and gain network access. This remediation process only works for systems with network access. Other models are using the TPM for secure boot like, bitlocker in windows. In TPM 1.2 the chip didn't offer any physical protection and defense was weak. In some vendor implementations the TPM 2.0 has physical protection at the die level that adds a tamper resistant mesh, this helps with hardware side channel attacks. The TPM can be used to valid specific application PCR values, but this needs to be implemented with custom software to generate and validate the values.

We have discussed the trusted boot process of using a TPM for establishing a root of trust during system bring up. We now move into a concept called Trusted Execution Environment. The ARM TrustedZone provides the memory management to be partitioned and restricted for secure applications. ARM has been working toward Security Services for DDS security plugins using ARM TrustedZone [5] [27]. Figure 10 shows a flow from normal to secure world where the securitylib (new name libddssec) communicates to the security libraries running in restricted memory space. Figure 10 shows the architecture Cortex -A (applications 32/64-bit architecture) and for Cortex -M (embedded 32-bit architecture) the TEE layer might be removed with direct firmware communications. The exception levels, EL0 -user space application, EL1 -privileged OS, EL2 – Hypervisor, and EL3 – firmware/security monitor. Cortex-A supports TrustZone with Memory Management Unit and the Cortex-M supports TrustZone with Memory Protection Units (MPU). Physical memory is divided into Normal or Secure by setting NS bit within the Translation Lookahead Buffer (TLB) for all system memory on an A architecture, where the MPU is programmed for different regions. Physical memory size on an M architecture is 4GBytes, where on the A architecture it can grow by adding RAM. Secure world has access to Normal, but not the reverse. A context switch is performed from normal world to secure by a calling an API. The M35P is an interesting chip with claim for anti-tamping protection against side channel attacks.

While the trusted execution environments vendors make claims that they are secure, and researchers continuously scrutinizing TEEs have discovered several vulnerabilities. There is a position paper that points out limited functionality within the secure region for, no mechanism to verify execution code, no defense within the secure region, no detection mechanism, and no mitigation when comprised [28]. Other work has been focused on side channels, power management and cache timing attacks on ARM processors with Trustzone [29] [30] [31]. Intel has its version called Software Guard Extensions (SGX), but this has been comprised leveraging the speculative execution bug [32]. Also, from a performance point of view, it was covered in this paper [3] that enabling security added latency, throughput and speed overhead to each of the transactions. By performing context switches between normal and secure world, this will surely add additional performance penalties for saving the state data to registers, startup, teardown and data validation process between the worlds., this is a consideration for real-time constraints. Since the secuirutylib work that ARM is working on is to enable the security plugins to live in the secure world, this would help mitigate against the OpenSSL dump routine during execution time but does not cover the data at rest or off-line attacks, for example a cold boot attack [33].

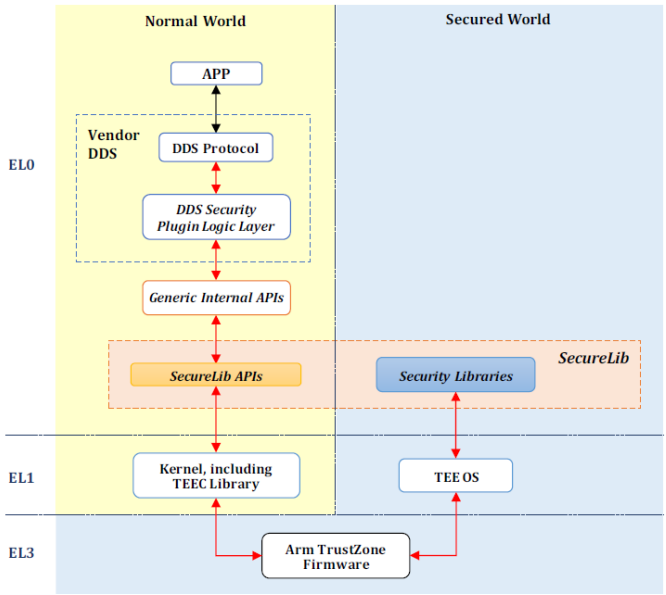

*Figure 10: Security Services for DDS security plugins using ARM TrustZone*

Our recommendation on Linux is to use IMA/EVM, since this technology has been up streamed and supported in the kernel [34]. IMA maintains a run time integrity list and is anchored to PCR 10 in the TPM. By extending the PCR 10 for each file listed in the policy, each measurement is aggregated into a value. This makes an attack difficult, since all the sequences and values must be known to re construct the result. Features of IMA include [35]:

    **Collect** – measure a file before it is accessed.

    **Store** – add the measurement to a kernel resident list and, if a hardware Trusted Platform Module (TPM) is present, extend the IMA PCR

    **Attest** – if present, use the TPM to sign the IMA PCR value, to allow a remote validation of the measurement list.

    **Appraise** – enforce local validation of a measurement against a "good" value stored in an extended attribute of the file.

    **Protect** – protect a file's security extended attributes (including appraisal hash) against off-line attack.

    **Audit** – audit the file hashes.

Table 3 shows a full software stack that includes the platform trusted services used to request quotes from the compute node, trusted software stack used as the interface for TPMs, IMA/EVM and the rest of the trusted boot as mentioned above [35].

*Table 3: Full stack of PTS, TSS, TPM and boot process using TCG specifications*

| Software Layer | Specification | Interface |
|---|---|---|
| Application | PTS | OpenPTS, TPM-Tools |
| Libraries | TSS | TrouSerS |
| Linux Kernel | TPM-2 | IMA, EVM, TPM Driver |
| Boot | BIOS | GRUB-IMA, TBOOT |
| Hardware | TPM | Software TPM |

By using the features of IMA, store and protect, the two templates below provide an example for each. Store provides a hash for the file to be generated, while protect adds a signature. An example of using an ima-ng template [35]:

PCR   template-hash   filedata-hash   filename-hint 10 91f34b5c671d73504b274a919661cf80dab1e127 ima-ng sha1:1801e1be3e65ef1eaa5c16617bec8f1274eaf6b3 boot_aggregate

Another example using an ima-sig template [35]:

PCR   template-hash   filedata-hash   filename-hint   file-signature 10 f63c10947347c71ff205ebfde5971009af27b0ba ima-sig sha256:6c118980083bccd259f069c2b3c3f3a2f5302d17a685409786564f4cf05 b3939 /usr/lib64/libgspell-1.so.1.0.0 0302046e6c10460100aa43a4b1136f45735669632ad ...

Additional information can be found in the kernel.org [36] about the IMA template management module. The protection provided by trusted boot and the IMA/EVM, this mechanism protects against offline attacks as well. Since IMA is a run time process files are being checked constantly, for changes. Having both OpenSSL and the property file under this type of control will mitigate several attacks, like the ones we demonstrated above.

## VI. CONCLUSION AND FUTURE WORK

We have presented two attack vectors related to ROS2/DDS security and have identified four different use case scenarios that are plausible. In each use case the adversary was able to obtain control of data or direct manipulation of the platform using the modified OpenSSL library and/or the configuration file with credential masquerading. In the first and second use cases, the data was either decrypted by possessing the correct keys or directly reading the plaintext data from the dump routines. In the third and fourth use cases either having the possession of the keys and/or credential masquerading could have achieved success by the adversary.

This paper presented the two attack vectors and compared technologies to help mitigate the risks. Even when files have been downloaded and checked with a hash, that one-time check does not provide the safeguards against ongoing threats. We believe that IMA/EVM can help mitigate against these threats, since it's a runtime and offline security set of features to protect files. The trusted boot does provide checks on the lower levels of the software stack to enable the OS to boot with a root of trust mechanism. While the Securitylib running in ARM Trustzone is an interesting protection mechanism, it only accounts for two of the five DDS security plugins. The need for a holistic security solution still needs to be considered for robotic architectures, since the new movement is toward autonomous. Enabling a TPM within a robotic platform and how attestation will be performed are new concepts that need to be extended beyond the traditional mechanism of trusted boot.

Future work, we believe that research is still at the early stages with consideration to using TPMs in robotic systems and how it's managed, validation of attestation performed by who and what remediation behavior will be put in place. New policies related to not just ethics, but security should be considered for robotic platforms. These are not IoT devices like some industry vendors are claiming and that a one solution can't be for all. Robots are moving into a cognitive learning phase where limited data is needed to learn and evolve within their environments. This is quite different from an IoT device.
.

## ACKNOWLEDGMENT

We would like to thank Real-Time Innovation (RTI) in supporting our research with early access product and support.